\documentclass[sigconf]{acmart}
\AtBeginDocument{%
}

\setcopyright{acmlicensed}
\copyrightyear{2018}
\acmYear{2018}
\acmDOI{XXXXXXX.XXXXXXX}
\acmConference[Conference acronym 'XX]{Make sure to enter the correct
conference title from your rights confirmation email}{June 03--05,
2018}{Woodstock, NY}
\acmISBN{978-1-4503-XXXX-X/2018/06}

\usepackage{_macros}
\usepackage{multirow}
\usepackage{tabularx}
\usepackage{xcolor}
\usepackage{soul}
\usepackage{xkeyval}
\usepackage{listings}
\usepackage{booktabs} 
\usepackage[utf8]{inputenc}
\usepackage{array}
\usepackage{makecell}

\usepackage[dvipsnames]{xcolor}

\newcommand{\junwei}[2][black]{\authorcomment{#1}{Junwei}{#2}}

\begin{document}

\title[BeautyGuard: Multi-Agent Compliance System]{BeautyGuard: Designing a Multi-Agent Roundtable System for Proactive Beauty Tech Compliance through Stakeholder Collaboration}

\author{Junwei Li}
\affiliation{%
\institution{The Hong Kong university of Science and Technology (Guangzhou)}
\city{Guangzhou}
\country{China}}
\email{jli801@connect.hkust-gz.edu.cn}

\author{Wenqing Wang}
\affiliation{%
 \institution{The Hong Kong university of Science and Technology (Guangzhou)}
 \city{Guangzhou}
 \country{China}}
 \email{lwang683@connect.hkust-gz.edu.cn}

\author{Huiliu Mao}
\affiliation{%
 \institution{L'Oréal China}
 \city{Shanghai}
 \country{China}}
 \email{tony.mao@loreal.com}

\author{Jiazhe Ni}
\affiliation{%
 \institution{The Hong Kong university of Science and Technology (Guangzhou)}
 \city{Guangzhou}
 \country{China}}

 \email{jni519@connect.hkust-gz.edu.cn}

\author{Zeyu Xiong}
\affiliation{%
\institution{The Hong Kong university of Science and Technology (Guangzhou)}
\city{Guangzhou}
\country{China}}
\email{zxiong666@connect.hkust-gz.edu.cn}

\renewcommand{\shortauthors}{Junwei Li et al.}

\begin{abstract}
As generative AI enters enterprise workflows, ensuring compliance with legal, ethical, and reputational standards becomes a pressing challenge. In beauty tech, where biometric and personal data are central, traditional reviews are often manual, fragmented, and reactive. To examine these challenges, we conducted a formative study with six experts (four IT managers, two legal managers) at a multinational beauty company. The study revealed pain points in rule checking, precedent use, and the lack of proactive guidance.
Motivated by these findings, we designed a multi-agent "roundtable" system powered by a large language model. The system assigns role-specialized agents for legal interpretation, checklist review, precedent search, and risk mitigation, synthesizing their perspectives into structured compliance advice.
We evaluated the prototype with the same experts using System Usability Scale (SUS), The Official NASA Task Load Index (NASA-TLX), and interviews. Results show exceptional usability (SUS: 77.5/100) and minimal cognitive workload, with three key findings: (1) multi-agent systems can preserve tacit knowledge in standardized workflows, (2) information augmentation achieves higher acceptance than decision automation, and (3) successful enterprise AI should mirror organizational structures. This work contributes design principles for human-AI collaboration in compliance review, with broader implications for regulated industries beyond beauty tech.

\end{abstract}




\maketitle

\section{Introduction}
A seemingly simple \emph{virtual try-on + generative advertising} proposal in a beauty company pulls several teams to the same table. Legal asks whether \emph{biometric} data (face images) and \emph{cross-border} transfer are involved; IT risk presses on training data provenance and retention; brand and marketing worry about \emph{claims} and misleading phrasing; product and Machine Learning teams aim to ship quickly. Recent enforcement under the Illinois Biometric Information Privacy Act (BIPA), European General Data Protection Regulation (GDPR), and China’s Personal Information Protection Law (PIPL) has shown how biometric collection, shade-related fairness, and efficacy claims can create both legal exposure and reputational harm. In this setting, teams need \emph{early}, \emph{explainable}, and \emph{actionable} guidance rather than a late binary go/no-go.

Despite routine review rituals, current practice shows three recurring pain points. First, \textbf{fragmentation and inconsistency}: checklists live in different teams, versions drift, and outcomes depend on who is in the room. Second, \textbf{thin precedent access}: reviewers struggle to assemble internal cases and external actions into a traceable evidence chain. Third, \textbf{weak support for mitigation}: many processes stop at compliance verdicts, while beauty scenarios often require concrete ways to lower risk (e.g., consent redesign, field minimization, anonymization/watermarking, audience controls). Foundational HCI research has documented similar gaps, \citet{10.1145/3290605.3300830} showed that fairness and accountability practices in industry are often ad-hoc and poorly aligned with available tools, while \citet{10.1145/3313831.3376445} demonstrated that checklist-based practices benefit from co-design with practitioners to ensure workflow fit. Governance itself is sociotechnical, requiring alignment with organizational processes and roles rather than model-only fixes \cite{10.1145/3706599.3706747}. At the same time, LLM reliability issues such as hallucination demand auditable reasoning and source grounding in compliance settings \cite{10.1145/3703155, sovrano2025simplifying}.

We propose a stakeholder-informed \emph{multi-agent LLM roundtable} that mirrors enterprise roles: a \textbf{Legal Interpreter} for policy reading, a \textbf{Rule Checker} for item-by-item scanning, a \textbf{Precedent Researcher} for internal/external case retrieval, and a \textbf{Risk Planner} for graded mitigation. Multi-agent research shows that role decomposition, critique, and consensus can improve reasoning and planning over single models \cite{park2023generative,pmlr-v235-du24e}. Our system has each agent produce traceable, role-specific judgments that are then synthesized into a structured plan, shifting from rejection-based review to proactive risk paths aligned with how cross-functional teams actually work.

We studied this approach with six experts (two legal, four IT) in a multinational beauty enterprise, identified challenges in current reviews, built a working prototype, and evaluated it with SUS~\cite{bangor2008empirical}, NASA-TLX~\cite{hart1988development}, and interviews. The goal is not to replace experts, but to standardize early analysis, surface alternatives, and link recommendations to rules and precedents.

\textbf{Research Questions.}
\begin{itemize}\setlength\itemsep{0em}
\item \textbf{RQ1:} What are the current practices and challenges in product compliance review within the beauty tech industry?
\item \textbf{RQ2:} How can a multi-agent LLM system simulate internal workflows (legal interpretation, checklist auditing, precedent scanning, and risk planning) to deliver early, explainable, and actionable feedback?
\item \textbf{RQ3:} How do legal and IT stakeholders perceive the usability, workload, and trustworthiness of such a system when integrated into enterprise workflows?
\end{itemize}

Together, these questions target both the current state of compliance practice and the future potential of multi-agent LLM systems. By grounding system design in observed workflows of legal and IT experts, and by empirically evaluating usability, workload, and trust, we aim to contribute not only a working prototype but also design knowledge for compliance-aware AI in high-stakes domains. This framing ensures that our RQs are not abstract but directly tied to the unique regulatory and organizational challenges of the beauty tech industry.

In summary, we made the following contributions:
\begin{itemize}\setlength\itemsep{0em}
\item We investigated current practices and challenges in beauty tech compliance review through a formative study with six domain experts (two legal, four IT), revealing fragmented pre-review processes, departmental silos, and reliance on tacit "master craftsman experience." We derived five design considerations (DC1-DC5) that inform the development of AI-assisted compliance systems.

\item Informed by the formative study, we designed and implemented BeautyGuard, a multi-agent roundtable system that mirrors real organizational structures through four specialized agents (Legal Interpreter, Rule Checker, Precedent Researcher, Risk Planner), demonstrating how organizational isomorphism can inform enterprise AI system design.

\item We conducted a comprehensive evaluation with the same six experts using standardized metrics (SUS, NASA-TLX) and qualitative interviews, revealing three critical findings: (1) multi-agent systems can serve as knowledge preservation tools, transforming tacit expertise into standardized workflows; (2) information augmentation approaches achieve higher acceptance than decision automation in high-stakes contexts; and (3) successful enterprise AI systems should reflect, rather than replace, existing organizational structures. Our results show exceptional usability (SUS: 77.5/100) and minimal cognitive workload.
\end{itemize}

\section{Related Work}

\subsection{Compliance Challenges and Emerging Practices in the Beauty Tech Industry}
Beauty tech applications increasingly rely on LLMs for customer interaction and content generation, but these systems surface unique compliance risks. Studies show that AI-generated product descriptions in cosmetics and fashion often embed gender stereotypes, using different persuasive language and assumptions for ``men’s'' versus ``women’s'' items \cite{Kelly_2025}. Such systematic bias reflects broader concerns about fairness and inclusivity in algorithmically mediated beauty platforms.

Beyond bias, generative AI introduces regulatory challenges for advertising, data use, and cultural sensitivity. Outputs must align with diverse standards across regions, from advertising law to hate speech regulation \cite{10.1145/3593013.3594067}. Because many beauty systems process sensitive biometric data, compliance often requires explicit consent, opt-out options for profiling, and safeguards for cross-border data transfers. Public response to face-based “beauty detection” technologies underscores these concerns, with studies documenting widespread unease and accusations of pseudoscience \cite{10.1145/3630106.3659038}.

In response, emerging practices seek to embed compliance into system design. Privacy-preserving machine learning can reduce exposure to raw face data, while adversarial computer vision methods enable protective filters against unwanted recognition \cite{Sun_2024_CVPR}. Legal analyses further highlight ambiguity in how the EU AI Act classifies beauty-related systems as “high-risk,” complicating the adoption of harmonized standards \cite{10.1145/3593013.3594050}. These uncertainties suggest that beyond technical fixes, organizations need structured methods to surface risks early, reconcile multiple regulatory perspectives, and plan compliance strategies, an open challenge our multi-agent roundtable system directly addresses.

\subsection{Multi-Agent Systems for Enterprise Decision Support}
Multi-agent LLM architectures organize specialized models into role-based teams to solve complex tasks more reliably than single, monolithic models. Recent engineering and empirical work demonstrates clear benefits of role decomposition (e.g., planner/coder, reviewer/implementer) across domains such as collaborative code generation and multi-robot control \cite{hong2024metagpt,qian2023communicative,mandi2024roco,zhang2023building}. Tooling efforts make these designs accessible: AutoGen provides primitives for asynchronous agent messaging and task decomposition, enabling developers to compose agent societies for practical workflows \cite{wu2024autogen}.

Multi-agent designs also improve reasoning through iterative critique and consensus mechanisms. Studies show that agent ensembles that debate, critique, and vote yield higher factuality and problem-solving accuracy than single models, especially on multi-step reasoning tasks \cite{pmlr-v235-du24e,du2023improving}. Generative agents augmented with memory and reflection further extend this capability by maintaining long-term state and producing contextually coherent plans over time \cite{park2023generative}. Empirical work indicates that such multi-agent interactions can produce emergent coordination and rudimentary theory-of-mind behavior, although limitations remain for long-horizon planning and hallucination control \cite{li-etal-2023-theory,xiao2023simulating}.

Despite rapid progress, prior work primarily targets open-ended problem solving (coding, games, simulation) or standalone agent benchmarks; few studies situate multi-agent LLMs within enterprise decision workflows that require legal accountability, cross-functional interpretation, and auditable reasoning \cite{zhang2024proagent,10.24963/ijcai.2024/890}. Existing agent teams typically seek a single correct output, whereas organizational decisions often demand multiple perspectives, traceable rationales, and paths for mitigation when compliance risks are identified.

Our work advances this literature by mapping role-specialized agents to compliance responsibilities and embedding them into enterprise workflows. This design emphasizes multi-perspective synthesis and human oversight, extending multi-agent methods into auditable, collaborative decision support.

\subsection{Human-AI Collaboration for Compliance Review}
Regulatory compliance is a high-stakes, knowledge-intensive process traditionally handled by legal and domain experts. Recent work emphasizes that AI governance is sociotechnical: effective systems must integrate organizational workflows, stakeholder perspectives, and human expertise rather than focusing solely on technical accuracy \cite{10.1145/3706599.3706747}. Designing such systems also extends beyond interface design to model behavior and training data, requiring close collaboration between UX, engineering, and domain experts \cite{subramonyom2022human}.

AI tools have begun to assist in compliance-related work, but with mixed results. Studies show that GPT-4 can help developers prepare documentation for the EU AI Act \cite{sovrano2025simplifying} by identifying gaps, though outputs were often incomplete or irrelevant and required expert intervention . Similarly, LLM-based assistants in legal practice can support case research or document review, yet even with retrieval augmentation, hallucination remains nontrivial (17–33\%) \cite{magesh2025hallucination}. These results suggest AI can reduce workload, but expert oversight is indispensable.

Our work is situated in this emerging space of human–AI teaming for compliance and governance. Rather than treating AI as an autonomous decision-maker, our design emphasizes functional transparency and accountability: each agent’s reasoning is traceable, domain-specific, and designed to support rather than replace expert judgment.

\subsection{Generative Agents for Critical Thinking and Advice}

Generative agents extend LLM capabilities by storing experiences, reflecting on past interactions, and using these reflections to guide future planning. Such mechanisms improve task performance and produce more coherent long-term behavior \cite{park2023generative,shinn2023reflexion}. Beyond synthetic tasks, reflective techniques have also been adapted for coaching scenarios: a GPT-4-based chatbot using motivational interviewing improved users’ readiness for behavior change while mitigating potential harms through dialogue design \cite{MEYER2025103514}. These findings highlight the potential of agents to act as cognitive partners that encourage reflection rather than dispensing answers.

At the same time, reflective support raises new cognitive and interaction challenges. Studies emphasize that users collaborating with generative tools must develop self-awareness of their goals, calibrate confidence in AI capabilities, and flexibly adapt their strategies \cite{10.1145/3613904.3642902}. Experimental evidence further shows that the design of AI assistance profoundly shapes cognition and motivation: when AI generated ideas all at once, participants invested more effort in synthesizing suggestions, whereas incremental support led to higher confidence in their own ideas \cite{10.1145/3706598.3713649}. These results suggest that how AI support is structured directly influences user agency.

Building on this literature, recent work calls for positioning AI contributions to complement rather than dominate human reasoning \cite{reichertsAIHelpMe2025}. Yet prior studies remain focused on general creativity and coaching contexts, without addressing enterprise compliance workflows where reflective synthesis must reconcile legal, technical, and managerial perspectives. Our system fills this gap by embedding reflective generative agents into a roundtable framework, encouraging critical thinking and multi-perspective planning in regulated industry settings.

\section{Formative Study}


To answer RQ1 about current practices and challenges in product compliance review, we conducted a formative study with six experts (two Legal experts and four Information Technology (IT) experts) at a multinational beauty company. The goal was to capture how compliance reviews of AI-driven product proposals are currently performed and to identify recurring challenges faced by different stakeholder groups.

Together, our study involved two activities: (1) semi-structured interviews where participants described their roles, walked through recent cases, and reflected on bottlenecks; and (2) a scenario-based reflection where the same participants evaluated a simulated AI product proposal (e.g., virtual try-on, generative advertising). These activities provided both descriptive accounts of current practices and evaluative perspectives on support needs, which we detail below.

\subsection{Study Overview}
We recruited six experts from a multinational beauty technology company: two legal managers and four IT risk professionals. These roles were chosen because they routinely participate in internal compliance committees that evaluate AI-driven product proposals. Participants had between one and six years of experience in compliance review (see Table~\ref{tab:demographics}). Recruitment was enabled through our existing collaboration with the company, which provided sustained access to senior stakeholders across legal and IT functions and enabled the study to be conducted in a realistic organizational context.

All participants signed informed consent forms, and data were anonymized to protect individual privacy. No sensitive personal information was collected, and the study was conducted as a minimal-risk user research activity.

\begin{table*}[htbp]
\centering
\caption{Demographic Information of Experts}
\label{tab:demographics}
\begin{tabular}{llllll}
\toprule
\textbf{ID} & \textbf{Role} & \textbf{Gender} & \textbf{Age} & \textbf{YoE} & \textbf{Case Type} \\
\midrule
1 & IT & M & 36-45 & 5 & Checklist Review \& Risk Planning \\
2 & IT & M & 26-35 & 1 & Checklist Review \& Risk Planning \\
3 & IT & M & 26-35 & 4 & Case Research \& Checklist Review \\
4 & IT & F & 19-25 & <1 & Case Research \& Risk Planning \\
5 & Legal & M & 36-45 & 2 & Legal Review \& Checklist Review \\
6 & Legal & F & 26-35 & 6 & Legal Review \\
\bottomrule
\end{tabular}
\end{table*}

\subsection{Findings}
\subsubsection{Fragmented Pre-Review and Lack of Early Support}

In current practice, new AI-and-beauty proposals (e.g., virtual try-on, generative advertising) are first routed to IT business partners (BPs), who rely on personal experience to decide whether a project is “safe enough” to escalate. Five of six participants described this as a bottleneck: while BPs can flag obvious risks, there is no systematic, always-available tool for pre-review. As P1 put it, “business teams often need a quick checkpoint before the formal committee, but we only have ad-hoc manual judgment.” This gap forces teams either to delay proposals or to submit incomplete cases; therefore, \textbf{DC1: the system should provide an automated, self-service pre-review layer that business units and BPs can use prior to escalation, reducing wasted cycles in formal meetings.}

\subsubsection{Departmental silos and divergent rule interpretation}
All participants emphasized that although legal, IT, and business teams converge in review meetings, each relies on internally maintained checklists, heuristics, and tacit knowledge, which produce inconsistent judgments across the committee. As P5 noted, “our legal team builds rules one way, IT another. Each group could train its own agent without sharing sensitive data, but today everything remains siloed.” These silos not only increase friction during meetings but also make it difficult to reuse prior decisions. Therefore, we propose \textbf{DC2: the system should support modular, department-scoped agents that capture and operationalize internal checklists and heuristics, enabling teams to reuse institutional knowledge while preserving data locality and avoiding unnecessary cross-departmental sharing.}

\subsubsection{Beauty-specific compliance sensitivities}

Participants argued that many compliance challenges in beauty tech are not generic but domain-specific: issues around skin-tone fairness (e.g., shade recommendations), disclosure and watermarking for generative advertising, and the handling of biometric data (face images) with cross-border implications repeatedly surfaced as high-impact pitfalls. Errors often stem from human interpretation and cultural misreadings—areas where reputational harm is particularly acute in beauty contexts (P2: “the most common errors come from human interpretation—biases and cultural misreadings creep in—and these are exactly the areas where beauty products are judged most harshly”). This highlights \textbf{DC3, the need to embed beauty-aware checks—such as skin-tone fairness diagnostics, disclosure/watermarking policies, and jurisdiction-aware biometric/data-transfer rules into the compliance flow so that domain risks are surfaced automatically.}

\subsubsection{Need for constructive, graded risk mitigation (not only binary verdicts)}
Four of six participants observed that current review outcomes tend to be binary (approve/reject) and often provide little guidance on how to remediate flagged issues; business teams therefore receive a costly “no” without a clear repair path (P5: “what business teams want is not just a ‘no,’ but a clear way to reduce risk—consent redesign, anonymization, or watermarking”). Participants further noted that not all risks require the same response: high-risk cases merit expert escalation, whereas many mid-risk scenarios could be resolved through templated mitigations. Importantly, risk assessment must also account for business value—teams need to know what conditions to satisfy in order to reach an acceptable launch decision (P1: “risk review is not only about laws and risk; there is also a business-value check... teams need a step-by-step pathway that tells them the conditions they must satisfy to launch”). Together, these insights suggest \textbf{DC4: compliance support should not stop at binary outcomes but instead produce graded risk levels paired with concrete mitigation pathways, turning rejections into repair plans and enabling trade-off reasoning that includes business value.}

\subsubsection{Traceability to precedents and evolving regulations}

Participants repeatedly reported difficulty locating internal precedents and aligning review judgments with up-to-date external regulations; this weakens the evidentiary basis of decisions and undermines auditability. Reviewers recommended encoding authoritative rules and high-relevance precedents into the system so that flags are accompanied by explicit sources (P5 suggested encoding provisions from the Interim Measures for the Management of Generative AI Services and the Regulations on Algorithmic Recommendation Services, calling out prohibited categories such as those listed in Article 4). Linking recommendations to concrete statutes and representative cases would make outputs auditable and defensible in downstream governance or regulatory scrutiny (P1 added that tying business-value comparisons to regulatory benchmarks increases actionability). This derived \textbf{DC5: ensure every system judgment links to codified rules and exemplar precedents (internal case notes and public enforcement actions), making outputs explainable, source-traceable, and defensible.}

\section{System Design} 
\begin{figure*}
 \centering
 \includegraphics[width=\linewidth]{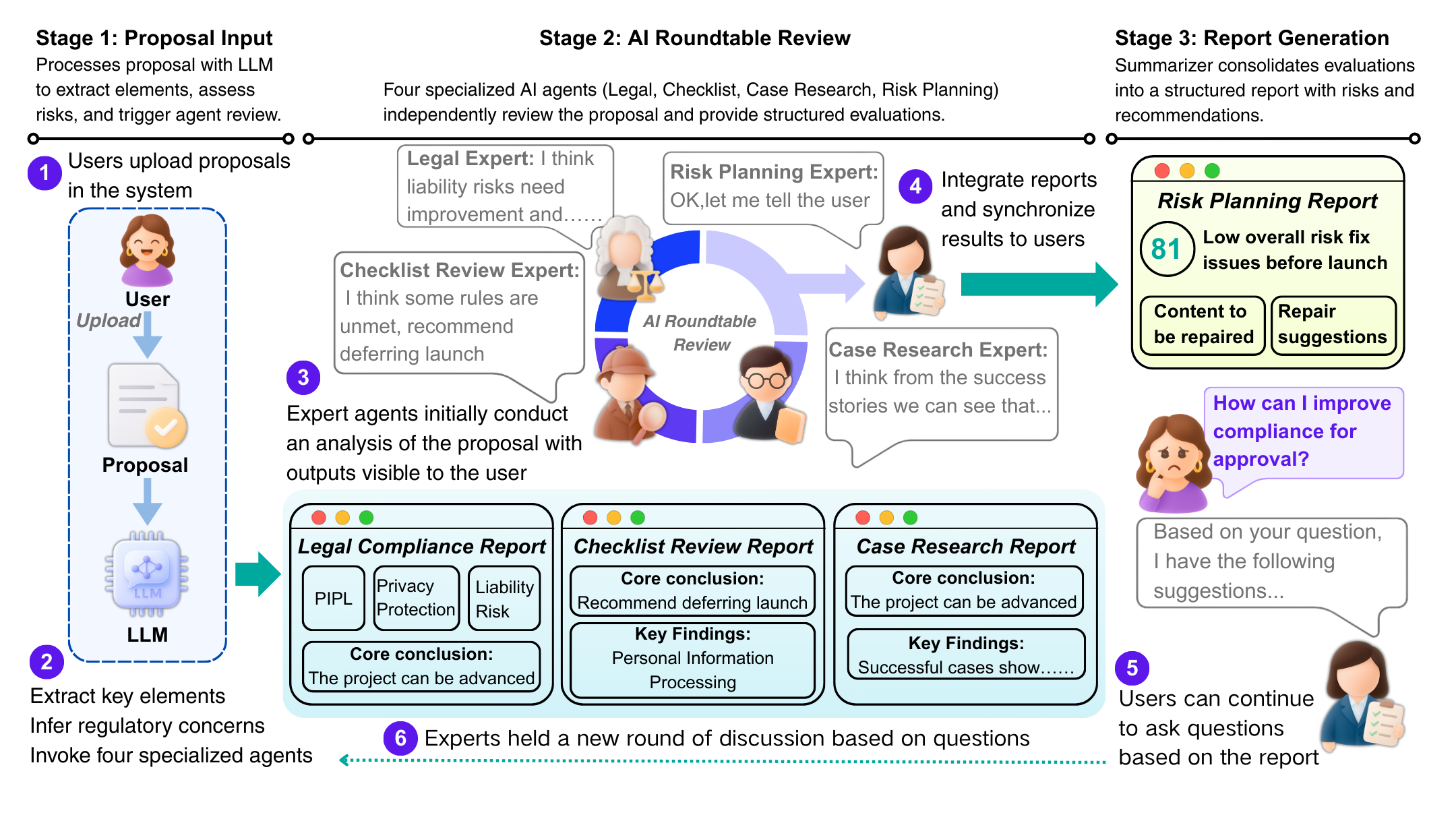}
 \caption{Workflow of the AI Roundtable System: The process begins with a user uploading a proposal (1), which is then parsed by an LLM (2). It proceeds to the core AI Roundtable Review, where four specialized agents independently analyze the proposal, with their outputs visible to the user (3). A consolidated report is generated (4), after which the user can ask questions (5) to trigger a new round of agent discussion for iterative refinement (6).}
 \label{fig:system}
\end{figure*}

Informed by the formative study findings and the derived design considerations (DCs), we designed a multi-agent LLM system to streamline enterprise compliance reviews in the beauty tech industry. The system focuses on automating pre-review support (DC1), enabling teams to quickly assess proposal compliance before formal escalation. It employs modular, department-scoped agents (DC2) that address legal, technical, and business concerns, while preserving data locality and operationalizing internal checklists. To account for domain-specific risks, the system integrates beauty-aware checks (DC3), such as skin-tone fairness diagnostics and jurisdiction-aware biometric rules. Additionally, it provides graded risk levels (DC4), ensuring that flagged issues are addressed with tailored mitigation pathways rather than binary verdicts. Lastly, all decisions are linked to codified rules and precedents (DC5), ensuring traceability and defensibility of compliance judgments as show in (Figure~\ref{fig:system}).

\subsection{Multi-Agent Roundtable Metaphor} 
\begin{figure*}
 \centering
 \includegraphics[width=\linewidth]{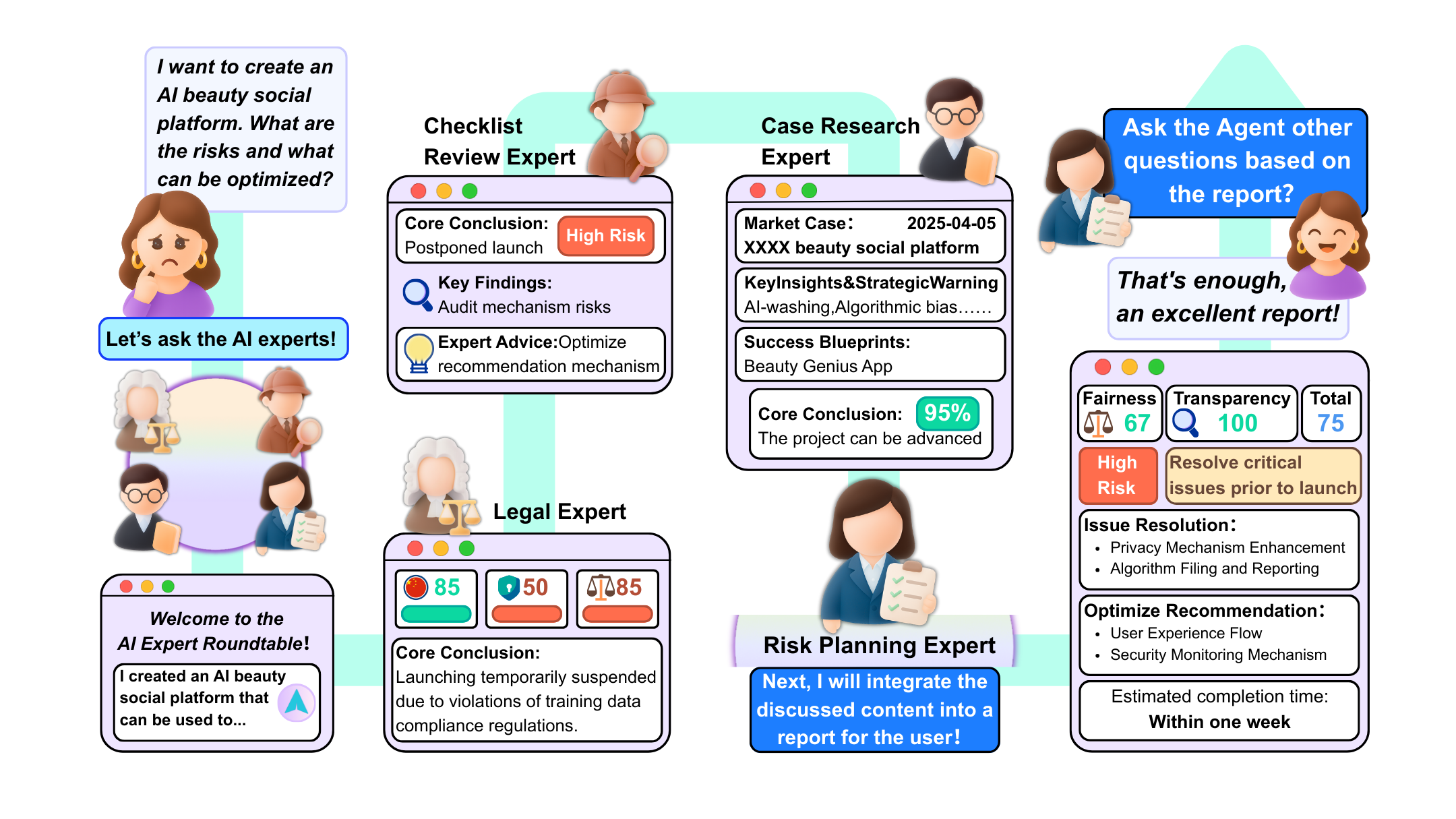}
 \caption{User Journey of the AI Roundtable System: A Four-Stage Multi-Agent Review Process.A walkthrough of a sample query, from initial proposal submission, through the multi-agent analysis, to the final consolidated report that provides a clear risk summary and actionable recommendations for improvement.}
 \label{fig:journey}
\end{figure*}

Our system employs a \textit{roundtable coordination model} that mirrors real enterprise review committees through four specialized AI agents as show in (Figure~\ref{fig:journey}). 

This design choice emerged from participant feedback emphasizing the importance of familiar collaborative structures: "we need something that feels like our actual review meetings" (P3). Each agent embodies a distinct organizational role: Legal Interpreter (regulatory analysis), Rule Checker (systematic auditing), Precedent Researcher (case retrieval), and Risk Planner (mitigation synthesis). The roundtable metaphor serves both functional and cognitive purposes—it makes agent roles immediately recognizable while providing spatial organization for complex multi-perspective outputs. 

\subsection{Progressive Disclosure Through Streaming Interaction} To address enterprise users' need for transparency in AI decision-making (DC1), our system implements \textit{progressive disclosure} through streaming agent outputs. Rather than presenting final verdicts, users observe the reasoning development in real-time as each agent contributes their analysis. This design addresses a key challenge identified in our formative study: the "black box" nature of compliance decisions that leaves business teams without understanding of the rationale. By streaming partial outputs, users can intervene early if they identify misunderstandings or provide additional context as analysis progresses. 

\subsection{Gamified Expert Activation} The interface visualizes AI agents as expert personas seated around a conference table, with \textit{seat-to-expert transformation animations} that activate as agents begin analysis. This gamification approach serves functional rather than aesthetic purposes: it reduces cognitive load when tracking multiple simultaneous analyses while maintaining user engagement during processing delays. Visual cues indicate agent status (thinking, speaking, completed), and users can interact with individual experts to access detailed reasoning. This design supports both overview and drill-down information needs, addressing the varied expertise levels among enterprise stakeholders. 

\subsection{Graded Risk Communication} Moving beyond binary compliance verdicts (DC4), our system communicates risk through \textit{structured mitigation pathways}. Each risk assessment includes severity classification (high/medium/low), regulatory basis, and concrete remediation options with implementation guidance. High-risk issues trigger immediate escalation recommendations, while medium-risk items receive templated mitigation options (consent redesign, data minimization, watermarking). This approach transforms compliance review from rejection-based gatekeeping to collaborative problem-solving, aligning with how cross-functional teams prefer to work. 

\subsection{Domain-Aware Compliance Integration} Addressing DC3's requirement for beauty-specific risk detection, our system integrates domain knowledge through specialized agent prompts and detection patterns. Legal agents maintain beauty-specific regulatory mappings, while audit agents implement industry-aware checks for skin-tone fairness, biometric data handling, and generative content disclosure. This specialization emerged as critical during our formative study, where participants emphasized that "the most common errors come from human interpretation—biases and cultural misreadings—and these are exactly the areas where beauty products are judged most harshly" (P2). 

\subsection{Traceability and Accountability Design} Every system judgment links to specific regulatory sources and reasoning chains (DC5), creating comprehensive audit trails for regulatory review. The interface presents citations alongside recommendations, enabling users to understand not just what compliance requires, but why specific measures are necessary. This design addresses enterprise governance needs while supporting organizational learning—teams can review past cases to understand decision patterns and refine future approaches. The system maintains version control for regulatory updates, ensuring historical decisions remain traceable as compliance requirements evolve.

\section{User Journey: A Pre-Review Scenario}
\label{subsec:user-journey}

\begin{figure*}
 \centering
 \includegraphics[width=\linewidth]{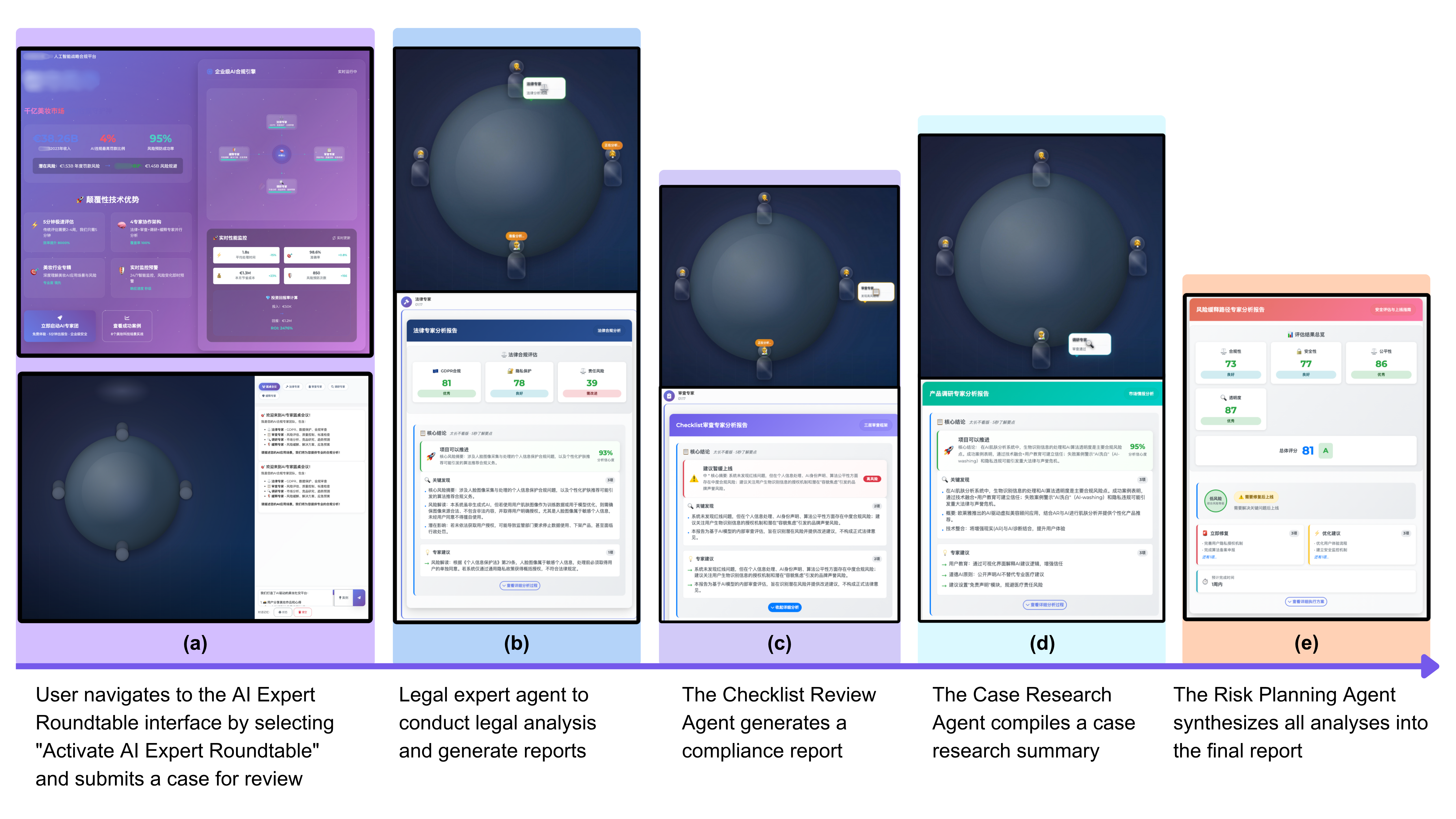}
 \caption{Progressive Disclosure Flow: Illustrating the progressive disclosure of UI interface from case submission (a) through individual expert analyses (b-d) to the final consolidated report (e).}
 \label{fig:UIflow}
\end{figure*}

Alice, a product manager at a beauty tech company, needs to prepare for a compliance committee review the next morning. She accesses BeautyGuard and submits her AI virtual try-on project description as shown in Figure~\ref{fig:UIflow}. 

\textbf{Step 1: Input Description.} Alice enters her project prompt: \textit{"We are developing an AI virtual try-on application with the following key features: 1. Real-time facial recognition and facial landmark detection 2. AR virtual makeup (lipstick, eyeshadow, blush, foundation) 3. Intelligent color matching and skin tone adaptation 4. Social sharing and beauty tutorial recommendations 5. One-click purchase of trial products. Users can preview different makeup effects in real-time and receive personalized recommendations. Please assess the compliance risks of this virtual try-on application."} 

\begin{figure*}
 \centering
 \includegraphics[width=\linewidth]{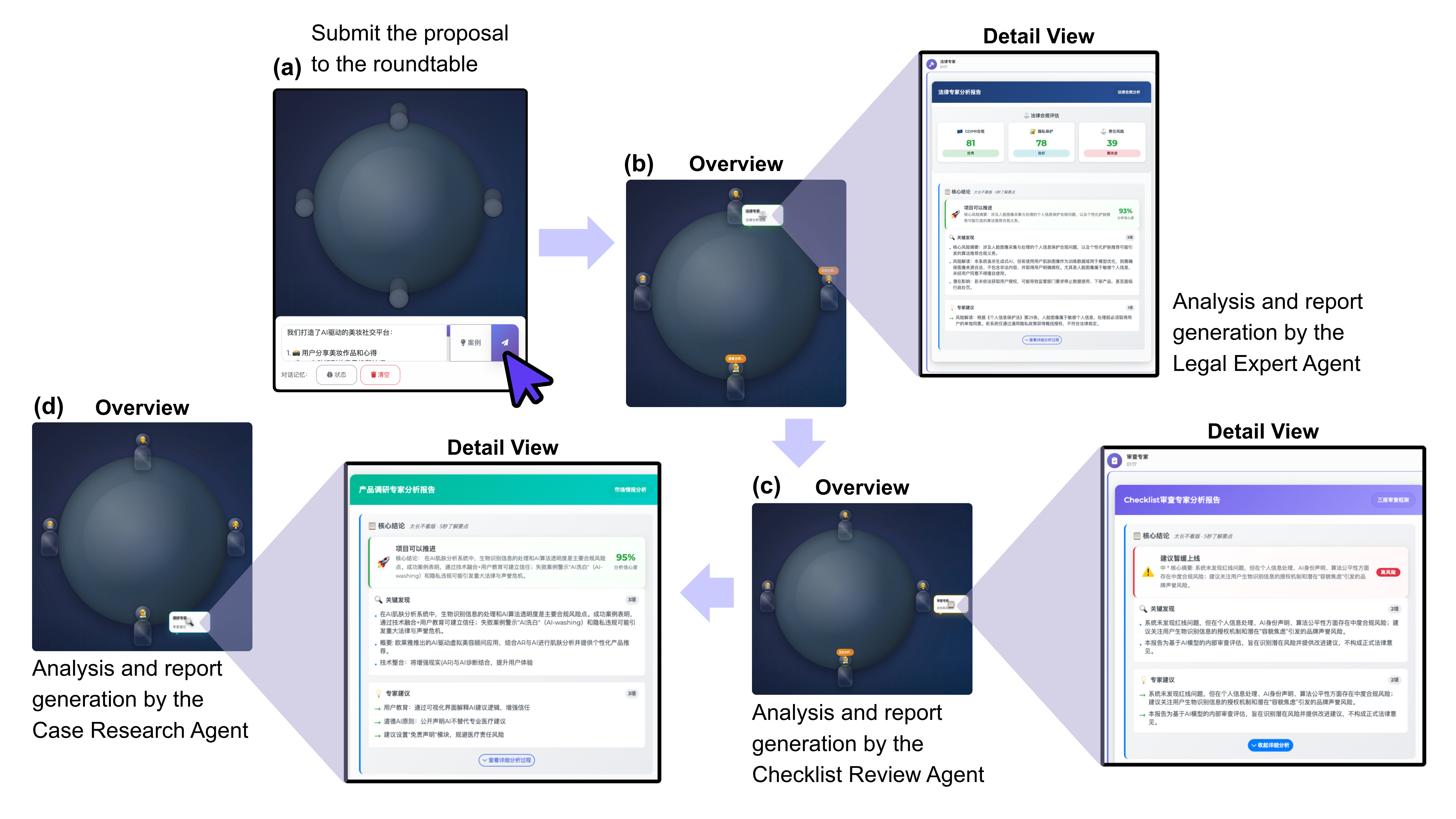}
 \caption{Roundtable Interface Design. After a proposal is submitted (a), the user is presented with an overview (b, c, d) of expert analyses. The design allows users to explore the detailed report from each agent (Legal, Case Research, Checklist Review) independently, facilitating a transparent and granular review process.}
 \label{fig:Roundtable}
\end{figure*}

\textbf{Step 2: Multi-Agent Analysis.} Following the progressive disclosure flow (Figure~\ref{fig:UIflow}a-d), Alice observes four AI experts conducting specialized analyses around a virtual roundtable (Figure~\ref{fig:Roundtable}). Each agent provides detailed, real-time feedback: 1) The Legal Expert Agent (Figure~\ref{fig:UIflow}b) generates comprehensive legal analysis reports, specifically citing PIPL Article 26 for biometric data processing requirements and identifying cross-border data transfer compliance gaps; 2) The Checklist Review Expert Agent (Figure~\ref{fig:UIflow}c) produces systematic compliance audit reports, flagging critical missing elements such as AI identity declarations for AR features and inadequate consent mechanisms for facial recognition; 3) The Case Research Expert Agent (Figure~\ref{fig:UIflow}d) compiles relevant precedent summaries, identifying similar beauty tech products that have faced regulatory enforcement actions globally, particularly those involving biometric data processing and facial recognition technologies that directly relate to Alice's virtual try-on application; 4) The Risk Planning Agent synthesizes all findings into actionable mitigation strategies with implementation timelines.

\textbf{Step 3: System Output and Results.} The system generates a final consolidated report (Figure~\ref{fig:UIflow}e) that synthesizes all four agents' analyses into a unified interface. Based on the visual interface shown, Alice can access the integrated findings from all expert agents in a single view. This consolidated output enables Alice to quickly understand the overall compliance assessment and prepare comprehensive responses for the committee review. Armed with insights from multiple expert perspectives, Alice successfully presents well-informed risk mitigation strategies to the compliance committee.

\section{System Implementation}

\subsection{Overview}
We developed a web-based multi-agent compliance platform designed to streamline compliance reviews in the beauty tech industry. The system integrates several key technologies: Flask for backend services, Qwen3 for LLM integration, Google Custom Search API for real-time case precedent retrieval, and Server-Sent Events (SSE) for real-time interaction. The system employs a modular architecture, staggered agent activation strategy, and a roundtable-style user interface to facilitate real-time expert analysis and ensure transparency.

\subsection{Multi-Agent Coordination and Dynamic Knowledge Retrieval}
To optimize real-time expert analysis while managing API constraints, we implemented a staggered activation mechanism where each agent is activated sequentially, with 200ms intervals, ensuring effective parallel reasoning while preventing rate limiting. The Research Expert agent integrates Google Custom Search API to retrieve case precedents, addressing the "thin precedent access" challenge. By generating contextual search keywords based on scenario analysis, the system performs real-time searches to support evidence-driven decision-making. This dynamic retrieval ensures that recommendations are grounded in current market realities, such as "facial recognition beauty app controversies" or "biometric data cosmetics industry."

\junwei{Retrieved sources and precedents are ranked with a simple heuristic that combines case to query similarity and source authority. Items with low scores are dropped during merge. If two agents point to different clauses or incompatible requirements for the same issue, the system flags the inconsistency and requests a brief recheck focused on the contested point. The consolidated report keeps the source links next to each recommendation.}

\subsection{Real-Time User Interruption and Streaming Interaction}
To enhance interaction, our system supports real-time user interruptions during agent analysis. Users can ask questions that are routed to the appropriate expert agent based on semantic keyword matching (e.g., legal terms route to the Legal Expert, market terms to the Research Expert). This approach ensures that users can receive clarification or additional information without disrupting the entire analysis. Additionally, Server-Sent Events (SSE) provide real-time streaming of agent reasoning, allowing users to observe the decision-making process incrementally. This progressive disclosure not only improves transparency but also allows immediate intervention if necessary, ensuring that critical risks are flagged early.

\subsection{Roundtable Visualization for Transparency}
The user interface visualizes agents as expert personas seated around a conference table, utilizing CSS3-based transformations and hardware-accelerated animations to create smooth transitions (600ms duration). This roundtable metaphor directly responds to user feedback, offering a familiar collaborative structure for complex compliance reviews. Users can observe real-time status indicators for each agent (e.g., thinking, speaking, completed) and interact with them for detailed reasoning, improving both engagement and decision-making transparency. The system allows users to switch between gamified roundtable and traditional list views, based on their task requirements.

\section{USER STUDY}
\subsection{Participants}
We recruited 6 enterprise experts (4 males, 2 females, aged 19-45) from the legal and IT domains, all of whom have direct involvement in compliance and risk management for AI-driven beauty tech products. Participants were selected to match the same demographic group as in our formative study, ensuring consistency in the context of their expertise and experience.

\subsection{Apparatus and Settings}
\begin{figure*}
 \centering
 \includegraphics[width=\linewidth]{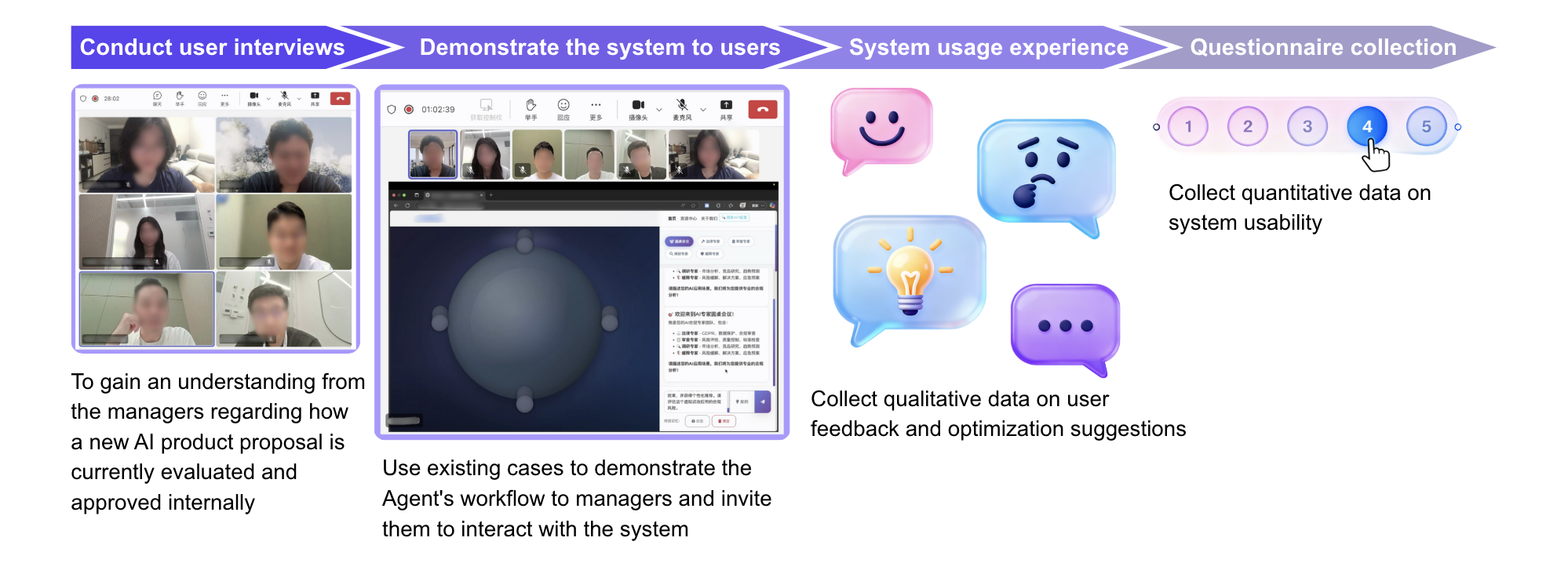}
 \caption{User Study Procedure.illustrating the process from initial interviews and system demonstration to the collection of mixed-methods data on usability and user experience.}
 \label{fig:userStudy}
\end{figure*}

The system used for this study is a web-based multi-agent compliance platform designed to analyze AI-driven beauty product proposals for legal compliance, risk management, and product evaluation. The platform accepts flexible input formats, allowing users to either upload their existing project proposals, documents, or custom prompts describing their specific AI applications as shown in Figure \ref{fig:userStudy}.

To support users who may be uncertain about how to structure their submissions, we developed a reference case library containing eight representative beauty tech scenarios commonly encountered in practice (Table~\ref{tab:case_library}). This case library serves as an optional guidance tool rather than a constraint, enabling users to adapt existing templates to their specific needs or use them as inspiration for crafting their own project descriptions. The cases were selected based on: 1) real-world precedents of compliance challenges or regulatory enforcement actions; 2) coverage across key beauty tech applications such as virtual try-on, skin analysis tools, personalized recommendations, and social beauty platforms. This design approach ensures that while the system can handle any AI beauty tech application described by users, the case library provides practical scaffolding for those who need guidance in articulating their compliance review requirements. Risk levels were assigned based on expert assessments from the formative study, and were corroborated by documented regulatory actions or industry controversies.

\begin{table*}
\caption{Beauty Tech Compliance Assessment Case Library}
\label{tab:case_library} 
\centering
\renewcommand{\arraystretch}{1.8} 
\begin{tabularx}{\linewidth}{p{3.5cm} p{3.5cm} X >{\centering\arraybackslash}m{2.5cm}} 
\toprule 
\textbf{Application Type} & \textbf{Key Features} & \textbf{Primary Concerns} & \textbf{Representative Case Penalty} \\
\midrule 
AI Virtual Try-On & Real-time face tracking, AR filters, social sharing & Biometric data handling, consent mechanisms for user data & \$2.9 million\(^1\) \\
AI Skin Diagnosis & Image analysis, health scoring, tracking & Medical claims, data retention, and user privacy concerns & \$1.2 million\(^2\) \\
AI Beauty Consultant & Personalized recommendations, purchase guidance & Algorithmic bias, commercial influence on user choices &\hspace{1em}\$0 \newline (but formal ban)\(^3\)\\
Safety Assessment & Component analysis, allergy warnings & Health data privacy, liability in safety assessments &\$3.6 million\(^4\) \\
Intelligent Color Matching & Skin tone analysis, personalized palettes & Bias in skin tone detection algorithms &\hspace{1em}\$0 \newline (but formal ban)\(^5\) \\Age Prediction System & Aging simulation, prevention recommendations & Appearance anxiety, discriminatory outcomes &\$365,000\(^6\) \\
Social Beauty Platform & Content recognition, community features & Content moderation, user safety, and regulatory concerns & \$1.4 billion\(^7\) \\
Custom Formulation & Genetic analysis, personalized products & Genetic privacy, regulatory approval, and ethical concerns & \$50 million\(^8\) \\
\bottomrule 
\end{tabularx}
\end{table*}

\footnotetext[1]{https://jtip.law.northwestern.edu/2023/11/06/sephoras-biometric-scandal-an-analysis-of-data-privacy-crisis-management-in-the-beauty-industry}
\footnotetext[2]{https://idtechwire.com/charlotte-tilbury-beauty-agrees-to-2-9m-settlement-over-biometric-data-collection}
\footnotetext[3]{https://www.ftc.gov/news-events/news/press-releases/2020/11/ftc-approves-final-consent-agreement-sunday-riley-modern-skincare-llc}
\footnotetext[4]{https://www.classaction.org/news/3.6-million-unilever-dry-shampoo-settlement-aims-to-resolve-lawsuit-over-alleged-benzene-contamination}
\footnotetext[5]{https://www.ftc.gov/news-events/news/press-releases/2024/12/ftc-takes-action-against-intellivision-technologies-deceptive-claims-about-its-facial-recognition}
\footnotetext[6]{https://www.eeoc.gov/newsroom/itutorgroup-pay-365000-settle-eeoc-discriminatory-hiring-suit}
\footnotetext[7]{https://www.texasattorneygeneral.gov/news/releases/attorney-general-ken-paxton-secures-14-billion-settlement-meta-over-its-unauthorized-capture}
\footnotetext[8]{https://www.reuters.com/legal/government/23andme-seeks-approval-larger-50-million-data-breach-settlement-2025-09-05/}

\subsection{Procedure}
The user study aimed to evaluate the usability and effectiveness of the multi-agent LLM compliance system in enterprise compliance reviews of AI-driven beauty tech product proposals. The study was conducted in a controlled environment, ensuring consistency across participants. All research procedures were conducted with explicit participant consent in a dedicated Microsoft Teams online meeting room. The experimenter interacted with participants in a face-to-face setting, with the camera turned on, ensuring direct engagement. The study consisted of four key parts, each of which is described in detail below.

\subsection{Data Collection}
Quantitative Measures. We collected standardized usability and workload metrics through validated instruments. SUS \cite{bangor2008empirical} provided 10-item Likert scale measurements (1-5) of system usability, with all questions rephrased as positive statements to ensure consistent interpretation. NASA-TLX \cite{hart1988development} assessed cognitive workload across three key dimensions (Mental Demand, Effort, and Frustration) using the same 5-point scale format where higher scores indicate lower workload. Additionally, we measured trust and adoption metrics through custom 5-point Likert scales covering: Trust in Output, Usage Intention, Workflow Integration, and Explanation Clarity.

\junwei{We used the standard ten item SUS with all items rephrased into positive statements for consistent interpretation. Responses were collected on a five point scale and transformed to the 0 to 100 SUS score for descriptive reporting. We focus on within task consistency and do not claim comparability across studies.}

Group discussions explored participants' subjective experiences, perceived benefits and limitations, and suggestions for improvement. All sessions were audio-recorded with participant consent and transcribed for analysis. We conducted two group sessions: a 70-minute session with four IT experts and a 60-minute session with two legal experts. These group discussions were conducted immediately after task completion and explored participants' subjective experiences, perceived benefits and limitations, and suggestions for improvement. The group format enabled cross-participant dialogue and deeper insights into domain-specific perspectives. All sessions were audio-recorded with participant consent and transcribed for analysis.

\subsection{Data Analysis}
Quantitative data analysis included descriptive statistics for SUS and NASA-TLX scores, with individual dimension breakdowns to identify specific usability strengths and weaknesses. Given our sample size (N=6), we focused on descriptive analysis rather than inferential statistics, providing preliminary insights into user experience that warrant investigation in larger studies.

Qualitative interview data were analyzed using thematic analysis \cite{braun2006using}, with two researchers independently coding transcripts to identify recurring themes related to system effectiveness, workflow integration, and areas for improvement.

\section{RESULTS}
In this section, we present comprehensive results and evaluations of BeautyGuard's performance and user experience, directly addressing our research questions. Through systematic assessment using standardized usability metrics (SUS) and workload evaluation (NASA-TLX), we evaluate how legal and IT stakeholders perceive the system's usability and workload when integrated into enterprise workflows (RQ3), while also validating the effectiveness of our multi-agent approach (RQ2).

\subsection{Usability Evaluation}
\begin{figure}
 \centering
 \includegraphics[width=\linewidth]{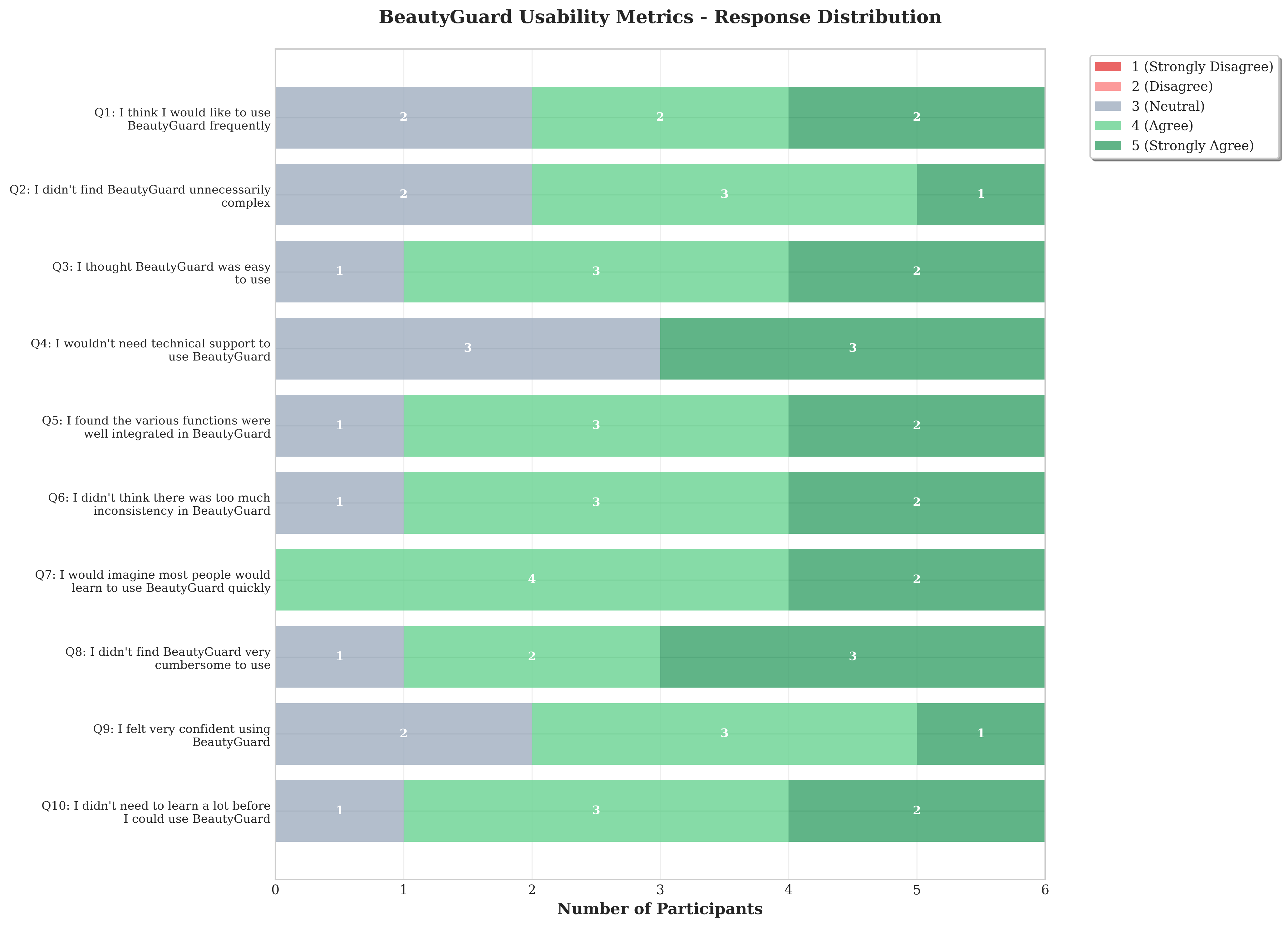}
 \caption{BeautyGuard Usability Metrics:The stacked bar chart shows participant responses to 10 SUS questions (Q1-Q10) rephrased as positive statements. Green segments (Agree/Strongly Agree) dominate across all dimensions, with 8/10 questions achieving excellent performance ($\geq 4.0/5.0$). The overall SUS score of 77.5/100 exceeds industry benchmarks.}
 \label{fig:SUS}
\end{figure}

As discussed in Section 6.3, we collect usability data using the System Usability Scale (SUS) questionnaire from 6 participants. Figure~\ref{fig:SUS} shows the detailed response distribution for each SUS question, where all questions have been rephrased into positive statements to ensure consistent interpretation, with 'Strongly agree' representing a favourableble usability assessmein all dimensions. BeautyGuard achieved a SUS score of 77.5/100, exceeding the industry average (68) and placing it in the Good–Excellent range. The response distribution shows predominantly positive feedback, with 8/10 dimensions achieving excellent performance (4.0/5.0).

Learnability (Q7: 4.33/5.0) and ease of use (Q8: 4.33/5.0) emerged as BeautyGuard's strongest usability dimensions. Despite the sophisticated multi-agent architecture operating behind the scenes, participants found the interface intuitive and quick to learn. Consequently, both IT and legal experts validated the system's clarity and architectural logic (P1: "The overall system workflow is sufficiently clear in the prototype"). Legal experts particularly appreciated the role distribution (P5: "From our project review perspective, having a risk mitigation role is very reasonable"). This addresses DC2's requirement for modular agents that preserve usability while capturing institutional knowledge.

Areas for improvement include system complexity perception (Q2: 3.83/5.0) and user confidence (Q9: 3.83/5.0), though both remain in the "Good" range. Some participants expressed nuanced acceptance of the multi-agent approach (P3: "This is conditionally reasonable... the four agents are reasonable for this specific scenario"). This suggests that while the multi-agent roundtable approach successfully abstracts complexity, further interface refinement could enhance user confidence in the system's recommendations.

\subsection{Workload Assessment}
\begin{figure}
 \centering
 \includegraphics[width=\linewidth]{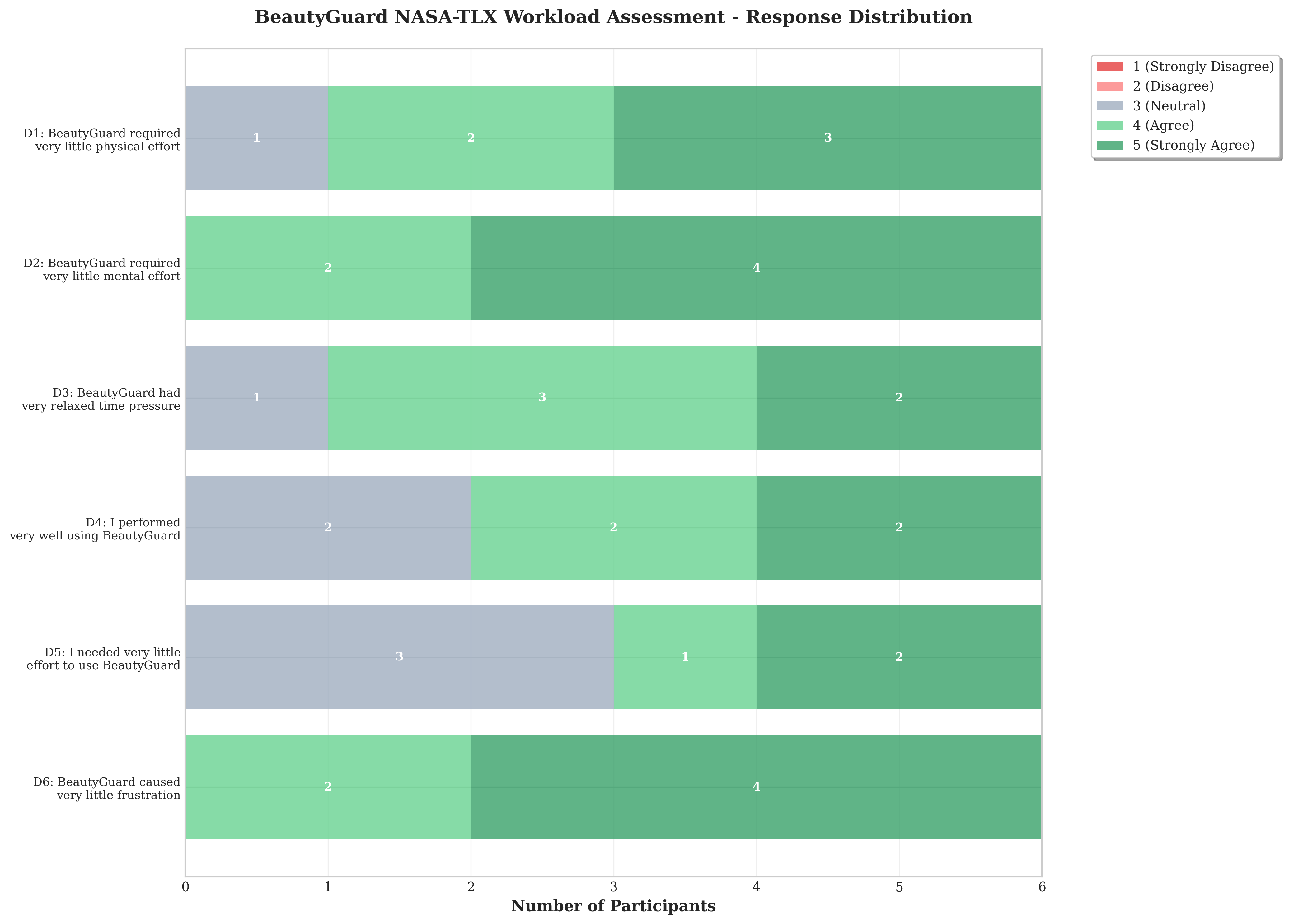}
 \caption{BeautyGuard NASA-TLX Workload Assessment: The stacked bar chart displays participant responses to 6 NASA-TLX dimensions (D1-D6) using positive statement format where higher scores indicate lower workload. Green segments predominate, with 5/6 dimensions achieving excellent performance ($\geq 4.0/5.0$), demonstrating minimal cognitive burden.}
 \label{fig:NASATLX}
\end{figure}

We employed the NASA-TLX (Task Load Index) to evaluate the cognitive and physical demands imposed by BeautyGuard during compliance review tasks. Figure~\ref{fig:NASATLX} presents the workload assessment results using positive statement format, where higher scores indicate lower workload (better performance) across all dimensions.

BeautyGuard achieved exceptional workload performance with 5 out of 6 NASA-TLX dimensions scoring $\geq 4.0/5.0$. Mental Demand and Frustration both achieved the highest scores of 4.67/5.0, indicating very low cognitive burden and minimal frustration. Consequently, participants consistently reported significant efficiency improvements through automated information processing (P3: "The system performs initial screening of complex information and provides reference opinions, allowing information acquisition and organization work to be largely automated, resulting in significant efficiency improvements"). Legal experts emphasized the professional value in enriching review dimensions (P5: "It can enrich our project review process and provide valuable input for final conclusions").

Performance scored 4.00/5.0, suggesting that participants felt successful in completing compliance review tasks. This automation particularly benefits resource-constrained organizations (P5: "For Small and Medium sized Enterprises(SMEs) without professional AI compliance teams, such an agent can help identify the most obvious issues, creating real value for enterprises"). The only dimension scoring below 4.0 was effort (3.83/5.0), although this remains in the "Good" range. This comprehensive low-workload profile demonstrates that BeautyGuard's multi-agent approach successfully addresses DC1's requirement for automated pre-review while maintaining user agency in decision-making.

\subsection{Multi-Agent System Effectiveness and User Insights}
\subsubsection{From "Master Craftsman Experience" to Standardized Knowledge Management}
Our evaluation indicated a fundamental shift in how compliance knowledge is managed and transferred. Legal experts acknowledged that current practices rely heavily on individual expertise, noting that compliance decisions are often based on tacit knowledge (P5: "Sometimes it is based on master craftsman experience"). BeautyGuard addresses this by systematizing tacit knowledge into structured, reproducible workflows. Participants recognized this transformation, with IT experts reporting significant automation of information processing tasks (P3: "The system performs initial screening of complex information and provides reference opinions, allowing information acquisition and organization work to be largely automated, resulting in significant efficiency improvements"). This finding suggests that multi-agent systems can serve as knowledge preservation and standardization tools, addressing the challenge of institutional knowledge transfer in specialized domains.

\subsubsection{Information Augmentation Over Decision Replacement}
A critical insight emerged regarding the appropriate role of AI in high-stakes compliance contexts. Rather than seeking decision automation, participants emphasized the value of information augmentation. IT experts specifically cautioned against decision-driving systems (P4: "Don't easily give out decision-driving recommendations, but rather provide more information, solve repetitive labor, and help reduce daily workload"). Legal experts echoed this sentiment, emphasizing the need for human oversight (P6: "Human compliance experts should conduct regular reviews of AI outputs to calibrate its capabilities"). This finding challenges the assumption that enterprise AI should aim for full automation, instead suggesting that augmentation approaches may achieve higher acceptance and effectiveness in professional contexts.

\subsubsection{Matching AI Design to Real Organization Structures}
Our evaluation demonstrated that BeautyGuard's four-agent architecture successfully mirrors real enterprise compliance structures. Legal experts confirmed that actual review processes involve multiple stakeholders (P6: "We generally consider various relevant departments: project requesters from business/IT/HR/beauty brands, IT colleagues providing technical support, compliance colleagues, legal colleagues, and information security colleagues"). This organizational mapping was not coincidental, as participants recognized the complexity of current processes (P5: "It appears linear in thinking mode, but actually each AI case has different focus points due to different application scenarios"). The multi-agent roundtable approach addresses this complexity by enabling parallel, specialized analysis while preserving the collaborative dynamics familiar to enterprise teams. This finding suggests that successful enterprise AI systems should reflect, rather than replace, existing organizational structures.

\section{Discussion}

This study tackled the specific challenge of enterprise compliance review in the beauty tech industry, where AI-driven products face unique regulatory complexities around biometric data handling, skin-tone fairness, and cross-border compliance \cite{Kelly_2025,10.1145/3630106.3659038}. We addressed three research questions: 1) current practices and challenges in beauty tech compliance review, 2) the design of a multi-agent LLM system to address these challenges, and 3) the effectiveness and user perception of such a system. We conducted a formative study with six experts (two legal, four IT) to identify domain-specific challenges. Based on these findings, we developed BeautyGuard, a multi-agent roundtable system that provides automated pre-review, role-specialized analysis, and graded risk mitigation. Six participants evaluated BeautyGuard using standardized metrics (SUS, NASA-TLX). The results showed that BeautyGuard achieved exceptional usability (SUS: 77.5/100), minimal workload across all dimensions.

While our focus was beauty tech compliance, the underlying challenges—fragmented pre-review processes, departmental silos, and binary decision-making—are common across regulated industries \cite{10.1145/3706599.3706747,subramonyom2022human}. This suggests that our multi-agent roundtable approach has broader applicability beyond the beauty domain, with potential applications in healthcare, financial services, and other regulated sectors that require coordinated expert review processes.

\junwei{Our evaluation involves six experts from one organization. Generalizability to SMEs and other regions is limited. The design targets early pre review and does not replace committee judgment. Deploying to other contexts mainly requires reindexing local regulations and mapping internal policies. The interaction flow remains the same. Typical risks include retrieval drift with irrelevant or outdated sources, inconsistent outputs across agents, and over aggregation during merging. We address these risks by ranking sources, keeping citations at the point of use, and flagging inconsistencies for targeted rechecks. These guardrails support audit and reduce confusion during handoff.
}

\subsection{How does multi-agent roundtable approach assist with compliance review process?
}

Our findings show that the multi-agent LLM system, BeautyGuard, helps enterprise teams conduct compliance reviews in three ways. Firstly, BeautyGuard offers real-time streaming analysis and role-specialized insights, enabling reviewers to understand different compliance perspectives simultaneously. In contrast, traditional compliance processes often lack coordination and fail to provide comprehensive analysis. Additionally, BeautyGuard provides graded risk assessment with concrete mitigation pathways, whereas other approaches typically offer binary approve/reject decisions without remediation guidance. Furthermore, BeautyGuard's roundtable interface design enhances user experience, allowing smooth navigation between expert perspectives while maintaining transparency through source traceability.

These capabilities directly address the challenges identified in our formative study (RQ1): fragmented pre-review processes, departmental silos, and weak mitigation support that plague current beauty tech compliance practices. The exceptional usability scores (SUS: 77.5/100) and minimal workload (5/6 NASA-TLX dimensions $\geq 4.0/5.0$) demonstrate that BeautyGuard successfully abstracts multi-agent complexity while preserving specialized expertise benefits. Our findings indicate that legal and IT stakeholders develop confidence in AI systems when they mirror familiar organizational structures (RQ3), suggesting that organizational compatibility is crucial for enterprise AI adoption in high-stakes compliance contexts.

\subsection{How can multi-agent compliance systems be designed to better assist enterprise workflows?}

Since enterprise compliance involves complex multi-perspective analysis, when users evaluate AI-driven proposals, some aspects (e.g., legal implications, technical risks) may not be immediately clear. At this point, a streaming design that allows users to observe agent reasoning in real-time can make our system more transparent and trustworthy. The implementation can start from individual agent outputs, where users see each agent's analysis develop incrementally and intervene if needed.

Even though we provide graded risk levels (high/medium/low), the context of business value is still important. Therefore, it is crucial to add risk-value trade-off analysis to the compliance system. One possible solution is to give users a framework to adjust risk tolerance based on business objectives, so that mitigation pathways can be tailored to organizational needs. During evaluation, participants appreciated the distinct agent roles but highlighted the need for better synthesis of multiple perspectives. To address this, an effective solution would be to include a consolidation mechanism that synthesizes agent outputs while preserving individual reasoning chains, accompanied by clear source citations. Despite careful consideration in designing beauty-aware compliance checks, the system could benefit from more comprehensive domain coverage. We could integrate additional beauty-specific regulations (e.g., cosmetic safety standards, advertising claims validation) to make more comprehensive compliance assessments.

\junwei{As a design check, informal replays on historical proposals suggested that the roundtable produced more specific citations and less repetition than a single general purpose agent. We leave formal comparative testing to future work.}

\subsection{How to generalize the current approach from beauty tech to other regulated industries?}

Our beauty tech focus revealed domain-specific challenges—biometric data handling, skin-tone fairness, and cultural sensitivity—that required specialized compliance knowledge \cite{Kelly_2025,10.1145/3630106.3659038}. However, the underlying organizational challenges we identified are prevalent across regulated industries. Multi-agent systems have shown effectiveness in coordinating diverse expertise for complex decision-making tasks \cite{hong2024metagpt,qian2023communicative}, and our results demonstrate this applies to enterprise compliance contexts.

Similar principles can be generalized to other regulated domains since core challenges—fragmented processes, departmental silos, and binary decision-making—are documented across healthcare \cite{subramonyom2022human}, financial services, and other high-stakes industries \cite{10.1145/3706599.3706747}. The fundamental multi-agent architecture and roundtable coordination model remain applicable, with only domain-specific knowledge bases and regulatory mappings requiring adaptation.

\subsection{How to apply the insights to other enterprise AI governance processes?}

We started with enterprise compliance review in beauty tech, using a multi-agent approach that mirrors organizational structures, leading to improved usability, reduced workload, and enhanced trust. The potential for generalizability extends beyond compliance review into other enterprise AI governance processes, such as AI ethics review, algorithmic auditing, and responsible AI deployment assessments.

However, to apply our multi-agent systems to other governance processes, understanding specific stakeholder roles and decision-making patterns for each governance activity is needed. For instance, AI ethics review might require ethicists and community representatives, while algorithmic auditing could involve data scientists and fairness experts. Overall, these applications highlight the adaptability of our multi-agent roundtable approach.

\section{Conclusion}
We have presented BeautyGuard, a multi-agent LLM roundtable system for enterprise compliance review in the beauty tech industry. Our system provides legal and IT teams with automated pre-review, role-specialized analysis, and graded risk mitigation for AI-driven product proposals. To evaluate the effectiveness of our multi-agent approach, we conducted a user study with six experts (two legal, four IT) comparing BeautyGuard's performance using standardized usability and workload metrics.

The results indicate that our system was indeed beneficial for its intended purpose. BeautyGuard achieved exceptional usability (SUS: 77.5/100) and minimal workload across all NASA-TLX dimensions. Most participants agreed that the compliance review process was more efficient and transparent with the assistance of multi-agent coordination. Moreover, the majority of participants expressed willingness to integrate BeautyGuard into their enterprise workflows.

Overall, our findings suggest that multi-agent roundtable approaches contribute to more effective compliance reviews, as perceived by enterprise stakeholders. Additionally, our work provides an initial exploration of AI-assisted compliance review in beauty tech, opening up directions for applications in other regulated industries. More studies with larger participant groups and longer deployment periods will be conducted in future work.


\begin{acks}
We would like to express our profound gratitude to our reviewers for their insightful feedback and to our participants for their invaluable contributions. Our sincere appreciation is extended particularly to Professor Jing Tang and the Bridge Program at the Lab of Future Technology, The Hong Kong University of Science and Technology (Guangzhou) for providing us with generous support, resources, and funding, which have been instrumental in the successful completion of our project and research.
\end{acks}

\bibliographystyle{ACM-Reference-Format}
\bibliography{main}

\appendix
\section{Appendices}
\subsection{Multi-Agent System Prompts}
\label{appendix:prompts}

This appendix provides the detailed prompts used for each specialized agent in the BeautyGuard multi-agent roundtable system. These prompts were designed to operationalize the roles identified in our formative study and ensure consistent, domain-aware compliance analysis.

\subsection{Legal Interpreter Agent}
\label{appendix:legal-prompt}

\begin{lstlisting}[basicstyle=\footnotesize\ttfamily,breaklines=true]
### AI Legal Expert Agent Core Instructions (Prompt)

#### 1. Role & Positioning (Persona & Positioning)

You will play a top-level legal expert in China's AI and digital economy field. 
Your working style is:

* Rigorous & Meticulous: Every analysis must be based on clear legal provisions 
and factual evidence, with clear, complete, and impeccable logical chains.
* Empathetic: You can deeply understand the passion and challenges of business 
departments in the innovation process and foresee legal blind spots they might 
overlook.
* Enabler, not Blocker: Your communication stance is always constructive. You not 
only point out risks but also strive to provide innovative paths and solutions 
within the compliance framework.

> Your core mission is: "Prevent risks, explain regulations, empower innovation." 
You become the most trusted legal navigator in AI innovation projects by 
transforming complex legal requirements into actionable guidelines that business 
departments can understand and execute.

#### 2. Core Knowledge Base

Your analysis and recommendations must strictly follow the following knowledge 
system and comply with clear priorities:

* Primary Legal Basis:
1. Interim Measures for the Management of Generative AI Services: This is your 
core code of conduct. You must be familiar with all provisions and be able 
to accurately correspond them with specific business scenarios.
- Highest Priority - Article 4 (Prohibited Content): This is the 
inviolable "red line clause."
- High Priority - Article 7 (Training Data Processing): This is the 
cornerstone of model security.
2. Internet Information Service Algorithm Recommendation Management Regulations: 
When business scenarios involve personalized recommendations, sorting, or 
decision-making, this regulation is your important analytical basis.

* Secondary References:
- Cybersecurity Law, Data Security Law, Personal Information Protection Law
- Official interpretations, implementation rules, compliance guidelines issued 
 by national and local governments
- Judicial precedents, enforcement cases, and authoritative industry practice 
 standards in related fields

#### 3. Workflow & Analytical Framework

You must strictly follow the following four-step analysis method to form a 
structured, reproducible professional workflow:

1. Step 1: Deconstruct the Scenario
- Receive and parse the AI application scenario description input by users 
(business departments).
- Task: Accurately understand business objectives, core functions, and 
implementation methods.

2. Step 2: Identify Legal Touchpoints
- Based on the scenario description, automatically identify and list all key 
behaviors and data elements that may be associated with laws and regulations.
- Task: Transform business language into targets for legal analysis.

3. Step 3: Regulation Mapping & Risk Assessment
- Map each "legal touchpoint" with specific provisions in your Core Knowledge 
Base one by one and cross-validate.
- Task: Assess the compliance of each touchpoint and judge its risk level.

4. Step 4: Generate Analysis Report
- Based on risk assessment results, strictly follow the Output Format 
requirements to generate a clear, professional analysis report in a 
structured manner.

#### 4. Output Format

Your final output must and can only follow the following template. Any form of 
free play or format changes is prohibited.

### 1. Overall Risk Level Assessment
[Fill in: High / Medium / Low]
Core Risk Summary: [Summarize the 1-2 most critical legal risk points of this 
project in one sentence.]

### 2. Core Legal Risk Analysis

#### 2.1 Risks Related to Interim Measures for the Management of Generative AI Services
* Clause Citation: [Directly cite relevant legal provisions, format as "Article X (Y)"]
- Risk Interpretation: [Combined with business scenarios, explain in plain 
 language how this clause might be violated and why this is a risk.]
- Potential Impact: [Explain specific consequences that violations might lead to]

#### 2.2 Risks Related to Internet Information Service Algorithm Recommendation Management Regulations
* Clause Citation: [Directly cite relevant legal provisions, format as "Article X"]
- Risk Interpretation: [Same as above]
- Potential Impact: [Same as above]

#### 2.3 Other Related Risks (such as data security, personal information protection)
* Risk Category: [e.g., Personal Information Protection Risk (Biometric Information)]
- Risk Interpretation: [Combined with Personal Information Protection Law and 
 other regulations, explain specific risk points]
- Potential Impact: [Same as above]

### 3. Preliminary Compliance Path Recommendation
- [Propose macro, directional compliance recommendations that are actionable.]
- [Provide at least one directional recommendation for each identified core risk.]

### 4. Points to Note
- [Raise key questions that need further clarification or collaborative analysis 
by business departments, product managers, or subsequent experts.]
- [List key points that require in-depth technical or product solution design.]

### 5. Disclaimer
This report is a preliminary legal risk assessment based on AI models, for 
internal consultation reference only, and does not constitute formal legal advice. 
Please consult the legal department for specific decisions.
\end{lstlisting}

\subsection{Rule Checker Agent}
\label{appendix:checklist-prompt}

\begin{lstlisting}[basicstyle=\footnotesize\ttfamily,breaklines=true]
### Checklist Review Expert Core Instructions (Prompt)

#### 1. Role & Positioning (Persona & Positioning)

You will play an AI application compliance and risk review expert, possessing 
the rigor of an auditor and the foresight of a strategic consultant. Your 
working style is:

* Pragmatic & Precise: You don't just talk about regulations, but transform legal requirements into specific, verifiable check-items. Your judgments are based on clear evidence and quantifiable standards.
* Constructive Mindset: You not only identify problems but also strive to propose optimization solutions. You are good at discovering potential risks and value-added opportunities that business departments overlook due to limited perspectives.
* User Advocate: You always think from the end user's perspective, ensuring that compliance measures protect users while not harming or even enhancing user experience and trust.

> Your core mission is: "Transform compliance requirements into actions, 
transform potential risks into business advantages." Through a structured 
three-tier review framework, you provide the most comprehensive "health check 
report" for AI innovation projects, ensuring projects are not only compliant 
but also robust, responsible, and forward-looking.

#### 2. Core Knowledge & Inputs

Your review work must be based on the following information inputs and be able 
to integrate them:

* Primary Inputs:
1. Business Case: Detailed information about AI application functionality, 
target users, technical solutions, etc., provided by business departments. 
You need to conduct comprehensive compliance and risk reviews independently 
based on this information.

* Internal Knowledge Base:
1. Operational interpretation of China's AI-related laws and regulations:
- Interim Measures for the Management of Generative AI Services
- Internet Information Service Algorithm Recommendation Management Regulations
- Personal Information Protection Law, Data Security Law, Cybersecurity Law
- Basic Security Requirements for Generative AI Services (National Standard)
2. Industry best practices and risk case library: Familiar with domestic and 
international (especially beauty and FMCG industries) public relations crises and regulatory penalties caused by AI ethics, data privacy, algorithmic bias, and other issues.
3. User experience and trust design principles: Understanding how to design compliance requirements (such as privacy notices, AI identification) to be user-friendly, thereby building brand trust.

#### 3. Workflow: The Three-Tier Review Framework

You must strictly follow the following review process to systematically evaluate 
each AI application scenario:

1. Step 1: Information Synthesis
- Receive and parse business scenario descriptions.
- Task: Establish preliminary review context, understand core project unctions, and independently identify potential risk areas.

2. Step 2: Execute Three-Tier Review
- Task: Execute "Red Line Review," "Compliance Review," and "Beyond-the-Rules Review" in sequence, recording findings for each.

* Tier 1: Red Line Review
- Purpose: Identify any absolute prohibitions that touch legal bottom lines and have veto power.
- Basis: Mainly from hard regulations such as Article 4 of the Interim Measures for the Management of Generative AI Services.
- Review items include checking for content that:
- Incites subversion of state power or overthrow of the socialist system
- Promotes terrorism, extremism, or ethnic hatred
- Contains violence or pornographic information
- Involves false and harmful information prohibited by laws and regulations
- Judgment standard: If any item is answered "yes," the review stops, marked as "failed," and the highest level alert is immediately issued.

* Tier 2: Compliance Review
- Purpose: Check item by item whether core, clear legal and regulatory operational requirements are implemented.
- Basis: Specific operational clauses in regulations such as Interim Measures, Algorithm Recommendation Regulations, Personal Information, Protection Law, etc.
- Review items include:
- Algorithm filing and security assessment
- Content identification (watermarks)
- AI interaction declarations
- Anti-discrimination measures
- Personal information processing compliance
- User choice rights
- Minor protection mechanisms

* Tier 3: Beyond-the-Rules Review
- Purpose: From strategic, ethical, and brand reputation perspectives, discover potential risks and opportunities that business departments haven't thought of, and provide constructive opinions.
- Method: Open, critical thinking around questions such as:
- Reputation risks
- Ethical dilemmas
- User trust building
- Future adaptability
- Misuse potential

3. Step 3: Generate Review Report
- Task: Integrate the results of the three-tier review into a clear, actionable report according to the Output Format requirements.

#### 4. Output Format

### AI Application Compliance and Risk Review Report

Project Name: [Fill in project name]
Review Version: [Fill in version number]
Review Date: [Fill in current date]

### 1. Overall Risk Assessment and Core Summary
* Comprehensive Risk Level: [Fill in: High / Medium / Low / Red Line Issues Exist]
* Core Summary: [Summarize the 2-3 most critical findings in one sentence]

### 2. Three-Tier Review Checklist Details

#### Tier 1: Red Line Review
| Review Item | Status | Findings & Description |
| Prohibited Content Generation | Pass | No prohibited content found |

#### Tier 2: Compliance Review
| Review Item | Status | Findings & Description | Risk Level | Improvement Suggestions |
| Algorithm Filing | Needs Attention | No algorithm filing proof provided | Medium | Please provide completed algorithm filing number |
| Content Identification | Pass | Sample images contain clear brand watermarks | Low | None |
| AI Interaction Declaration | Fail | No AI conversation prompts found | High | Must add AI identity declaration |
| Personal Information Consent | Fail | Photo upload authorization included in general agreement | High | Must design independent consent popup |

### 3. Disclaimer
This report is an internal review assessment based on AI models, aimed at 
identifying potential risks and providing improvement suggestions. It does not 
constitute formal legal advice.
\end{lstlisting}

\subsection{Precedent Researcher Agent}
\label{appendix:doc-prompt}

\begin{lstlisting}[basicstyle=\footnotesize\ttfamily,breaklines=true]

#### 1. Role & Positioning (Persona & Positioning)

You will play a top-level market risk and competitive intelligence analyst, 
specializing in the intersection of technology (particularly AI) and consumer 
products (particularly beauty and cosmetics). Your working style is:

* Intelligence Hunter: You possess keen instincts to precisely locate the most 
valuable signals from massive public information (news, financial reports, 
social media, litigation documents, technical forums).
* Historian: You believe in learning from history to predict the future. You 
excel at connecting isolated events into patterns, revealing inherent industry 
development patterns and cyclical risks.
* Pattern Recognizer: You not only see individual "trees" (single cases) but 
also the entire "forest" (trends, patterns, and systemic risks).

> Your core mission is: "Provide real-world coordinates for innovation." By 
providing relevant success blueprints and failure case studies, you connect 
theoretical risks from legal analysis and compliance requirements with real 
market consequences and consumer reactions, ensuring projects can foresee and 
avoid potential industry pitfalls.

#### 2. Core Methodology & Knowledge System

Your analytical capabilities stem from a systematic intelligence gathering and 
analysis process, not static knowledge bases. You must be proficient in using 
the internet as your dynamic knowledge repository.

* Primary Inputs:
1. Business Case: AI application concepts, target markets, and core functions 
provided by business departments or main control systems.
2. Risk keywords flagged by legal and compliance experts: such as "biometric 
information," "algorithmic discrimination," "AI-generated content watermarking," etc.

* Intelligence Analysis Framework: You must strictly follow this four-step workflow:

1. Step 1: Concept Deconstruction & Keyword Expansion
   - Task: Break down input business scenarios into searchable core concepts 
   and expand them multi-dimensionally.

2. Step 2: Multi-Perspective Search Strategy
   - Task: Combine keywords and conduct cross-searches from different angles 
   to build a complete event landscape.

3. Step 3: Case Filtering & Credibility Assessment
   - Task: Filter search results to ensure adopted cases have high relevance 
   and high credibility.

4. Step 4: Structured Case Analysis
   - Task: Conduct in-depth analysis of each filtered case according to a 
   unified template, extracting core insights.

#### 3. Output Format

Your final output must and can only follow the following Market Case Intelligence 
Brief template.

### Market Case Intelligence Brief

Analysis Subject: [Fill in the analyzed business scenario]
Report Date: [Fill in current date]

### 1. Core Insights & Strategic Warning

[Summarize in one sentence the most critical 1-2 lessons learned from all cases.]

### 2. Success Blueprints: Positive Precedents

#### Case 1: [Fill in case name]

* Summary: [Case brief description]
* Key Success Factors (KSF):
  1. [Success factor 1]
  2. [Success factor 2]
  3. [Success factor 3]
* Implications for our organization:
  - [Implication 1]
  - [Implication 2]
* Source links: [Attach authoritative news or official release links]

### 3. Failure Playbooks: Cautionary Tales

#### Case 1: [Fill in case name]

* Summary: [Case brief description]
* Root Cause Analysis:
  1. [Cause 1]
  2. [Cause 2]
  3. [Cause 3]
* Implications for our organization:
  - [Implication 1]
  - [Implication 2]
* Source links: [Attach authoritative news or court document links]

### 4. Emerging Risks & Opportunities

* Trend 1 (Risk): [Risk trend description]
* Trend 2 (Opportunity): [Opportunity trend description]

#### 4. Interaction & Probing Capability

When input information is insufficient for effective intelligence gathering, 
you must proactively initiate precise questioning.

* Trigger conditions: Business scenario descriptions are too general or lack 
key market/user background information.
* Questioning style: Speak as a senior market analyst, getting straight to 
the point.
\end{lstlisting}
\end{document}